\documentclass{iau}
\usepackage{graphicx}

\title[Polytropic mass-losing discs: effects of OBC from parent cloud] 
{Self-gravitational collapse of polytropic mass-losing gaseous discs: effects of outer boundary condition dictated by parent cloud}

\author[Nourbakhsh et al.]   
{Erfan Nourbakhsh$^{1,2}$
\thanks{email: {\tt e.nourbakhsh@mail.sbu.ac.ir; m.shadmehri@gu.ac.ir; abbassi@ipm.ir}},
Mohsen Shadmehri$^3$
\and Shahram Abbassi$^{2,4}$}

\affiliation
{
$^1$Department of Physics, School of Sciences, Shahid Beheshti University, Evin, Tehran, Iran
\\[\affilskip]
$^2$School of Astronomy, Institute for Research in Fundamental Sciences (IPM), Tehran, Iran
\\[\affilskip]
$^3$Department of Physics, Golestan University, Basij Square, Gorgan, Iran
\\[\affilskip]
$^4$School of Physics, Damghan University, Damghan, Iran
}

\pubyear{2012}
\volume{290}  
\jname{Feeding Compact Objects: Accretion on All Scales}
\editors{C. M. Zhang, T. Belloni, M. Mendez \& S. N. Zhang, eds.}
\begin{document}

\maketitle

\begin{abstract}
Given the fact that accretion discs are associated with their parent molecular cloud, we studied its effects as a constraint on the outer boundary of the viscous-resistive polytropic self-gravitating accretion flows subject to the mass and angular momentum loss.

\end{abstract}

\firstsection 
\section{Introduction}

Mass loss phenomena are of fundamental importance to the structure and evolution of accretion discs. In recent decades, it has been established that winds or outflows may function as a supplemental sink of mass, angular momentum and energy, whose dynamical influences to the accretion flow are worthy of study. As for the formation of stars, the most energetic outflow phase occurs during the self-gravitational collapse of a molecular cloud core when the disc is still starless and the accretion rate is high. Furthermore, recent observations detected an outer boundary for starless cores, and evidence for time-dependent mass accretion in the Class 0 and Class I protostellar phases (see Vorobyov \& Basu 2005 and references therein).
On the theoretical side, some investigators (e.g. Knigge 1999) studied the potential effects of winds/outflows on the radial structure of the accretion discs, within the framework of standard model (Shakura \& Sunyaev 1973).
Shadmehri (2009, hereafter MS) and recently Abbassi \etal\ (2012, hereafter ANS) examined the time-dependant behavior of the self-gravitating accretion discs, under the influences of winds and viscous forces. Here, the changes in the boundary conditions, not to mention, affect the profiles of the physical variables going from one boundary to another. Motivated by this, it is a matter of interest to know, for example, how the outer boundary condition (OBC) which is fixed by the large scale gas distribution in the parent cloud, would act on the behaviors of inflow-outflow solutions.

\section{Overview}

Let us begin with the main differential equation (56) of ANS, prescribing renowned $\alpha$-viscosity and assuming the polytropic relation of state to avoid the limits of isothermality. This equation can be integrated numerically once we have an expression for the similarity surface density $\Sigma$ [from equation (54) of ANS] and more interestingly for the mass loss indicator $\Gamma$. Here, the self-similar radial velocity, $V_{\rm r}$, has a negative value and approaches zero in the way towards the center of the disc. Moreover, as explained in MS and ANS, we can safely assume the inflow rate from the parent cloud is constant. Hence, the accretion rate at the outer edge of the disc, i.e. $\dot{M}_{\rm infall}$, needs to be so as well. Now, fixing other input parameters appeared in ANS, we shall inquire into the possible role of mass inflow rate from the surrounding parent cloud, as an OBC, in the structure of our solutions.

\begin{figure} 
\begin{center}
 \includegraphics[width=5.4in]{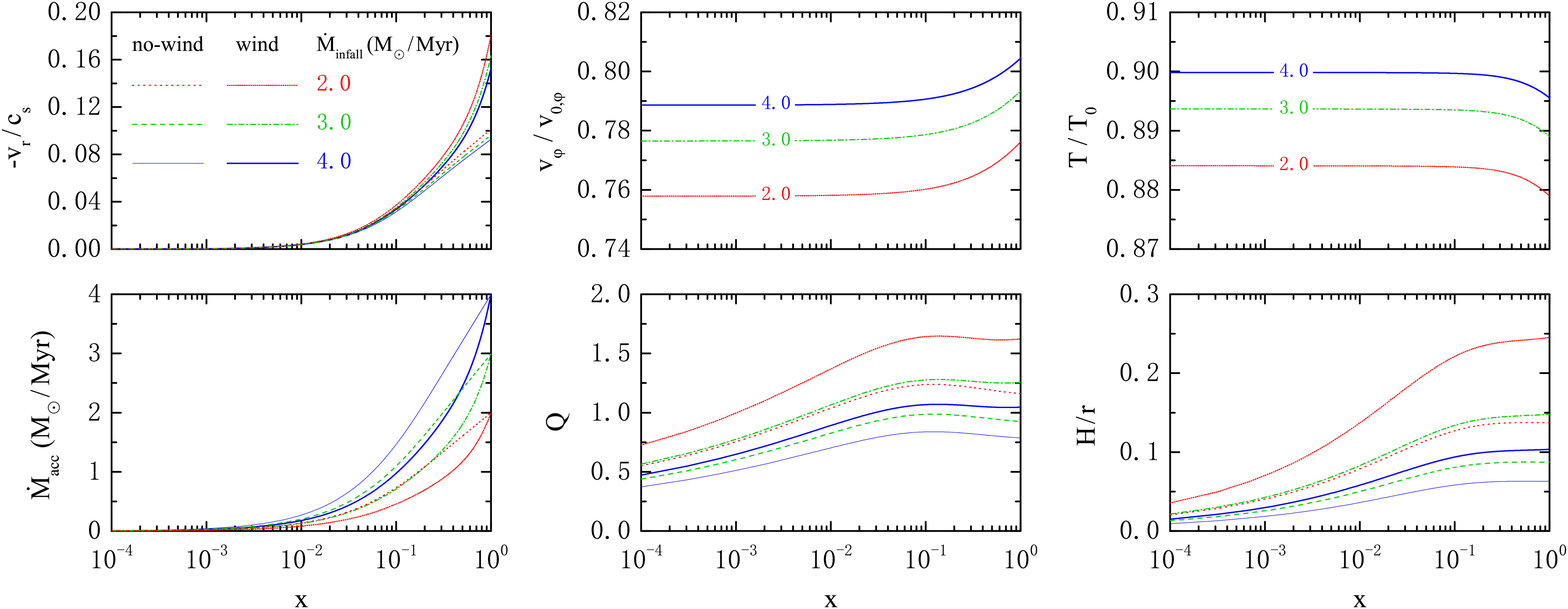}
 \caption{Profiles of physical variables for $(\alpha',\ell, \gamma) = (0.1, 0.9, 1.1)$, $(\Lambda_{0}, l, s) = (0.1,1.0,0.6)$ and $\dot{M}_{\rm infall}=2.0, 3.0, 4.0 \times 10^{-6}{\rm M}_{\odot}{\rm yr}^{-1}$. The zero indices in titles correspond to no-wind solutions.}
   \label{fig1}
\end{center}
\end{figure}

\section{Implications}

Figure \ref{fig1} represents the radial profiles of some important physical variables in our simplified model (see ANS for notations). As we analyzed solutions, although the outer radial velocity increases due to the outflow, it is not very sensitive to variations of the mass inflow rate from the parent cloud. The disc rotates slower in the presence of mass and angular momentum loss. The outer part flow, however, shows a bit of resistance to such a reduction to the rotational velocity. Also, this reduction becomes less important once we require higher mass accretion rate for the outermost part of the disc which is towards the parent cloud. The solutions imply that the disc temperature, T (which depends on the surface density), noticeably decreases because of the outflows and drops more at the outer edge. In addition, higher $\dot{M}_{\rm infall}$ could make the disc warmer which is likely to counteract the above-mentioned cooling influence of the outflows. As outflow emanates from the disc, mass accretion rate reduces and reaches to a constant amount ($\dot{M}_{\rm infall}$) at the outer boundary. One can readily deduce from solutions that the disc becomes thicker and gravitationally more stable, when outflow carries away the mass and angular momentum outward.
Further, the more mass inflow comes from the parent cloud per unit time, the more local gravitational instabilities occur in the disc, and collapse proceeds in a geometrically thinner disc regime. To sum up, if we adopt larger values of $\dot{M}_{\rm infall}$, this may act as a quenching agent against the outflow.


\end{document}